\documentstyle[prd,epsfig,aps,floats]{revtex}

\newcommand{\beq}{\begin{equation}}
\newcommand{\eeq}{\end{equation}}
\newcommand{\beqa}{\begin{eqnarray}}
\newcommand{\eeqa}{\end{eqnarray}}

\newcommand{\ie}{{\it i.e.}}
\newcommand{\eg}{{\it e.g.}}

\newcommand{\etal}{{\it et al.}}
\newcommand{\gev}{{\rm GeV}}

\newcommand{\M}{{\cal M}}

\newcommand{\R}{{\cal R}}

\newcommand{\jpsi}{J/\psi}

\newcommand{\eq}[1]{Eq.\ (\ref{#1})}

\newcommand{\ktr}{k_\perp}
\newcommand{\ptr}{p_\perp}

\newcommand{\llb}{{\lambda\bar\lambda}}
\newcommand{\ssb}{{\sigma\bar\sigma}}

\newcommand{\pvec}{\bbox{p}}

\newcommand{\qvec}{\bbox{q}}

\newcommand{\lvec}{\bbox{\ell}}

\newcommand{\lqcd}{\Lambda_{QCD}}

\newcommand{\qpair}{Q\bar Q}
\newcommand{\cpair}{c\bar c}

\newcommand{\PL}[3]{Phys.\ Lett.\ {{\bf#1}}, {#2} ({#3})}
\newcommand{\NP}[3]{Nucl.\ Phys.\ {{\bf#1}}, {#2} ({#3})}
\newcommand{\PR}[3]{Phys.\ Rev.\  {{\bf#1}}, {#2} ({#3})}
\newcommand{\PRL}[3]{Phys.\ Rev.\ Lett.\ {{\bf#1}}, {#2} ({#3})}
\newcommand{\ZP}[3]{Z. Phys.\ {{\bf#1}}, {#2} ({#3})}

\begin{document}

\twocolumn[\hsize\textwidth\columnwidth\hsize\csname @twocolumnfalse\endcsname
\title{%
\hbox to\hsize{\normalsize\hfil\rm NORDITA-98/55 HE}
\hbox to\hsize{\normalsize\hfil hep-ph/9809362}
\hbox to\hsize{\normalsize\hfil \protect\today}
\vskip 40pt
The Dynamics of Quarkonium Production\cite{byline1}}
\author{Paul Hoyer}
\address{Nordita\\
Blegdamsvej 17, DK--2100 Copenhagen, Denmark\\
www.nordita.dk}

\maketitle

\begin{abstract}
Quarkonium production is a sensitive probe of the dynamics of hard
scattering, which can `measure' the environment of the heavy quark pair
after it is created in a hard process. Quarkonium hadroproduction data
indicates that the produced pair experiences a secondary, hard interaction
and then ceases to interact at a stage when the pair is still compact
(compared to the size of the quarkonium wave function). These features
differ from scenarios mostly discussed so far. An approach which
relies on an early, perturbative rescattering of the pair with a comoving
color field can explain many observed features, including the polarization,
the $\chi_1/\chi_2$ ratio and the nuclear target $A$-dependence, which are
difficult to understand otherwise.
\end{abstract}
\vskip2.0pc]


\section{Introduction} \label{sec1}

The production of charm and bottom quarks provides a rich testing ground for
the dynamics of hard QCD processes. The detection of heavy quark(s) is
an unambiguous signal of a hard underlying process at the quark mass scale
$m_Q$, with no need for jet reconstruction. The heavy quark momentum is
(roughly) determined by that of the heavy hadron in which it is contained.
The quarks are colored and reinteract with other colored partons in the
final state, making heavy quark processes complementary to lepton pair
(Drell-Yan) production. Existing data on photo- and hadroproduction of open
and hidden charm and beauty production, at small and large transverse
momenta and on a variety of targets allows systematic studies of the
properties of hard processes. For a recent review of many aspects of open
heavy quark production see Ref. \cite{frixione}

In the standard QCD approach to hard inclusive processes the cross section is
written as a product of universal structure and fragmentation functions
times a process-specific PQCD subprocess. This is justified by the QCD
factorization theorem \cite{fact}, which applies when the sum over the final
hadronic states is well approximated by the partonic one. For heavy quarks
this is typically the case well above production threshold. In the
total heavy quark cross section the average transverse momentum of a heavy
quark is comparable to its mass, $\ptr \sim m_Q$. For large $m_Q$ many final
states are available, and the heavy quark hadronization is kinematically
well separated from the spectator partons of the colliding hadrons. The
total charm and beauty cross sections in both photo- and hadroproduction in
fact agree well with PQCD expectations \cite{frixione}.

A closer look at the production characteristics of charm quarks nevertheless
reveals some surprising features. One would expect the heavy quarks to be
produced nearly back-to-back in azimuthal angle, \ie, they should be
coplanar with the projectile. As seen from Fig. 1, this is rather well
satisfied for charm photoproduction, whereas for hadroproduction the measured
azimuthal distribution is nearly flat. To fit the hadroproduction data
requires a `primordial' transverse momentum of the partons in the colliding
hadrons as large as $\langle \ktr^2
\rangle \simeq 1\ \gev^2$, of the same order as the scale $m_c^2$ of the
hard process. 

The different characteristics of photo- and hadroproduction
suggests that the reason for the acoplanarity lies in gluon radiation of the
projectile parton. This radiation creates a `hot' environment for the heavy
quarks which may further increase their acoplanarity through secondary
interactions. Such interactions do not affect the factorization theorem for
observables (like the total cross section) which involve an inclusive sum
over the heavy quark momenta.

\begin{figure}[htb]
\center\leavevmode
\epsfxsize=8cm
\epsfbox{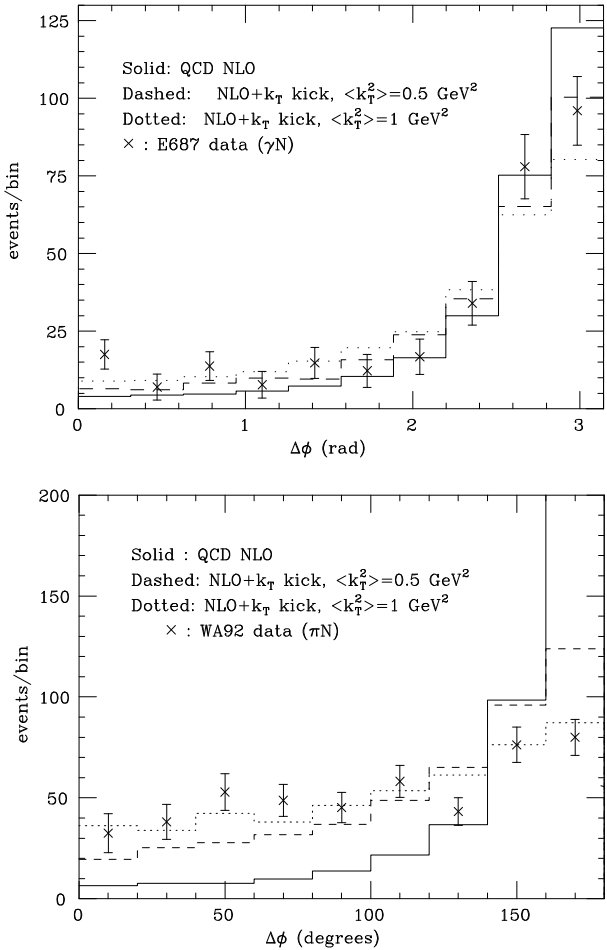}
\medskip
\caption{The relative azimuthal angle distribution of $D\bar D$ pairs in
photoproduction (upper figure) and hadroproduction (lower figure). Figures
taken from Ref. [1].}
\end{figure}

In quarkonium production the relative momentum of the heavy quarks
is severely constrained -- to a small fraction of $m_Q$ for a
non-relativistic bound state -- hence the QCD factorization theorem does
not apply. There is no {\em a priori} reason to believe that the
production cross section can be expressed as a product of universal
parton distributions and a PQCD subprocess. In particular, the hot
environment observed in $D\bar D$ production will be experienced also
by the heavy quarks that bind into quarkonia. This environment may well
affect the quarkonium cross section -- the bound state can be broken up
by considerably softer interactions than required to get the nearly
flat azimuthal distribution of Fig. 1.

Here I would like to focus on recent work done together with S. Peign\'e
\cite{hoypeg}. Based on properties of the quarkonium data we construct a
scenario for the production dynamics which is consistent with QCD. We also
discuss an explicit model which appears to explain several puzzling features
of the data.

\section{Hints from the quarkonium data} \label{sec2}

As noted above (Fig. 1), there is evidence from $D\bar D$ production
that hadroproduced heavy quarks are created together with a strong
color field. If this field is causing the `anomalies' of the
quarkonium cross section \cite{rev2,rev3} then one would expect no anomalies
in quarkonium photoproduction, in the fragmentation region of the
photon (which should be free of color fields). As a matter of fact, the Color
Singlet Model (CSM) provides a very good description of inelastic
$\jpsi$ photoproduction \cite{kzsz,h1photo,zeusphoto}.

The time scale of reinteraction effects between the heavy quarks
and the surrounding color field in hadroproduction can be surmised from the
ratio $\sigma(\psi')/\sigma_{dir}(\jpsi)$ of the $\psi'$ and the directly
(\ie, not through cascade decays) produced $\jpsi$. The measured ratio is
known to be independent of the nature of the projectile and target and of the
kinematics. Its value is furthermore consistent with the square of the ratio
of the wave functions at the origin \cite{vhbt,lourenco,gksssv}. This
suggests that the reinteractions occur while the heavy quark pair is still
in a compact configuration, compared to the spatial size of the quarkonium
wave function.

A further hint concerning the quarkonium production dynamics is provided by
the nuclear target $A$-dependence. The $\jpsi$ and
$\psi'$ cross sections both scale as $\sigma_A \simeq \sigma_N A^\alpha$,
with $\alpha = 0.92 \pm 0.01$ \cite{lourenco}. On the other hand, for open
charm $(D)$ production $\alpha = 1.02 \pm 0.03 \pm 0.02$ \cite{dprod}. This
difference in $A$-dependence is quite remarkable. At high energies the charm
pair remains in a compact $\sim 1/m_c$ configuration while it is inside the
nucleus. The uncertainty in its energy is thus so large that the pair
couples to both open and hidden charm channels. In other words, the
`decision' as to whether a given quark pair will materialize as a $\jpsi$,
$\psi'$ or $D\bar D$ final state can be made only after the pair has left
the nucleus. Apparently the only way of understanding their differing
$A$-dependence is then that {\em the formation of $\jpsi$ and $\psi'$, but
not of $D\bar D$, requires a further interaction with comoving partons}. If
the comovers are `swept away' by the nucleus a suppression of the quarkonium
cross section will result \cite{khasat}.

The transverse separation of such comovers from the quark pair can be
estimated from the effective `absorption' cross section needed to describe
the observed nuclear suppression. In a Glauber framework one finds
$\sigma_{abs}
\simeq 7.3 \pm 0.6$ mb \cite{klns} which (using $\sigma = \pi r^2$)
corresponds to a radius of about 0.5 fm. This is considerably bigger than
would be expected for a compact $c\bar c$ state ($r = 1/m_c \simeq 0.13$
fm), although it is still smaller than the typical QCD scale $1/\lqcd \simeq
1$ fm.

The anomalous effects in the cross section do not appear to vanish quickly
with the quark mass. The $\Upsilon(3S)$ cross section exceeds the CSM
prediction by an order of magnitude, and the nuclear suppression persists
(albeit at a reduced level). Whatever causes the anomalies seems to be of
`leading twist', in the sense that its effect vanishes slowly with $m_Q$.

Similarly, the anomalous effects observed in the total quarkonium cross
section do not decrease (if anything, they increase) with the transverse
momentum of the quarkonium. Thus, for $\ptr \gg m_c$ the CSM
underestimates the measured $\jpsi$ and $\chi_{c1}$ cross sections by more
than an order of magnitude, whereas the relative production rate of $\jpsi$
and $\psi'$ is close to expectations based on the wave function at the
origin. Since the anomalous effects are similar in the total and large
$\ptr$ cross sections it is reasonable to assume that they have the same
origin. The following discussion concerns the total cross section, but an
analogous scenario is applicable to the large $\ptr$ cross section.

\section{Rescattering from a comoving color field} \label{sec3}

How can we understand and describe a reinteraction mechanism, qualitatively
indicated by the data as discussed above, in terms of QCD? The loop diagram
shown in Fig. 2b may seem like a good candidate. The projectile gluon
radiates a gluon which reinteracts with the quark pair. However, on closer
inspection this diagram does not describe a two step process, with the
reinteraction clearly separated in time and space from the production of the
heavy quark pair. The spatial size of the loop is dictated by the
hard scale to be $1/m_c$. The reinteraction thus occurs simultaneously with
the creation of the heavy quarks, and the diagram must be viewed as a vertex
correction to the lowest order CSM diagram of Fig. 2a.

\begin{figure}[hbt]
\center\leavevmode
\epsfxsize=6.2cm
\epsfbox{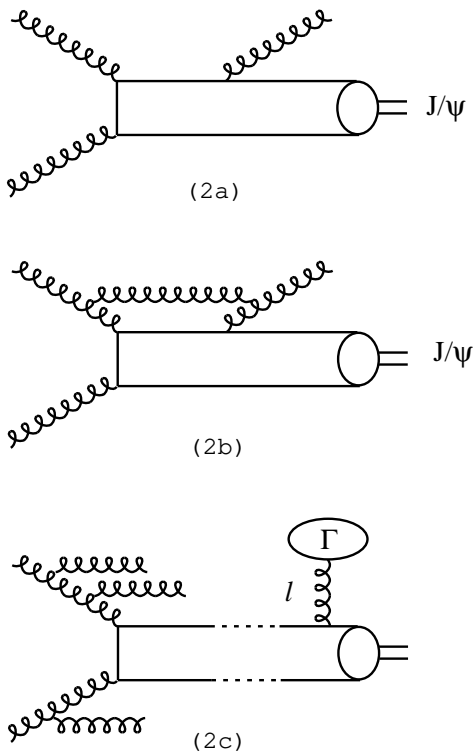}
\medskip
\caption{Basic processes for $\jpsi$ hadroproduction in the CSM (figures
(2a) and (2b)) and in the present scenario (figure (2c)).}
\end{figure}

A qualitatively different possibility is sketched in Fig. 2c. Here 
the quark pair is first created in a color octet state, together with the
color field radiated from the colliding gluons. This system propagates
for a time commensurate with the time-scale of emission of the radiated field
(which is long compared to the time $1/m_c$ in which the quark pair is
created). During the propagation the emitted field interacts
with itself and also with the quark pair. Most of these interactions are
relatively soft `monopole' ones, where the quark pair acts as a pointlike
colored object. The internal quantum numbers of the pair (\eg, the quark
spins) are thus unchanged. Eventually a harder dipole interaction (gluon
`$\ell$' in Fig. 2c) resolves the internal structure of the pair and changes
it into a color singlet (in this process the quark spins may flip). If
the color field has lost its phase coherence with the
quark pair it will act like a classical source.

The existence of a phase incoherent color field $\Gamma$ as in Fig. 2c
seems not to be excluded by any theoretical arguments. The qualitative
comparison of QED and QCD production processes in Fig. 3 may help to explain
how such a field could arise in hadroproduction, but still be
absent in photoproduction.

The upper half of Fig. 3 is a sketch of the QED process $e^+ e^- \to \mu^+
\mu^-$ in its CM, before and just after the creation of the muon
pair. The incoming electrons carry a radiation field, which is released when
the electrons annihilate. The right- and leftmoving fields do not interact
and thus simply pass each other, leaving the muon pair in an initially bare
state, without a surrounding (comoving) field.

In the QCD process $gg \to Q\bar Q$ (lower half of Fig. 3) the colliding
gluons again carry a radiation field. However, now the right- and leftmoving
fields interact as they pass each other, creating field components at lower
rapidities. Qualitatively, the situation seems analogous to the
creation of a rapidity plateau in hadron collisions. The final color field
could thus be evenly distributed in rapidity, as sketched in Fig. 3. The
external field $\Gamma$ of Fig. 2c would then correspond to the central
parts of this field, with low CM rapidity.

\begin{figure}[hbt]
\center\leavevmode
\epsfxsize=8cm
\epsfbox{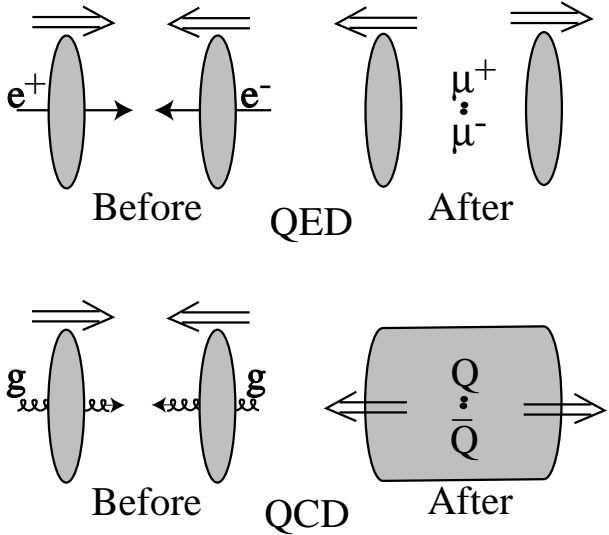}
\medskip
\caption{{\em Upper half:} Sketch of the $e^+ e^- \to \mu^+ \mu^-$ process
before (left) and just after (right) the creation of the muon pair. The
motion of the radiation field (grey) is indicated by double arrows.
{\em Lower half:} Analogous sketch of $gg \to Q\bar Q$. Interactions
between the colliding colored radiation fields create field components
which are comoving with the heavy quarks, \ie, which have low rapidity in the
$Q\bar Q$ rest frame.}
\end{figure}

In the subprocess $\gamma g \to Q\bar Q$ of heavy quark photoproduction,
only the gluon carries a radiation field. After the creation of the heavy
quarks this field then continues without interactions and emerges at high
rapidity in the final state. The bare heavy quarks are left without a
comoving field ($\Gamma=0$), much like in the QED example of Fig. 3. Hence
one would expect the CSM to work in photoproduction, as observed
experimentally.

\section{An explicit model of quarkonium hadroproduction} \label{sec4}

The scenario for quarkonium hadroproduction discussed in the previous
section is qualitative. Detailed predictions depend on the properties of
the comoving field $\Gamma$. Here I would like to briefly describe the
results of a model study, based on a number of simplifying assumptions
\cite{hoypeg}. In particular, we assumed that $\Gamma$ is weak enough so
that only a single `dipole' interaction needs to be considered, and that
$\Gamma$ is isotropically distributed in the rest frame of the $\qpair$. 

The scattering amplitude consists of three parts, and is calculated from the
diagram shown in Fig. 4. The first part is the $gg \to \qpair$ amplitude
$\Phi$, calculated using PQCD. The `monopole' interactions which occur after
the fusion process do not change the quantum numbers of the quark pair
(except for its octet color component) and thus need not be explicitly taken
into account. However, in the time interval between the creation of the pair
and its reinteraction with $\Gamma$ the spatial size of the quark pair
increases by some factor
$\rho$, which is a parameter of our model. 

\begin{figure}[htb]
\center\leavevmode
\epsfxsize=8cm
\epsfbox{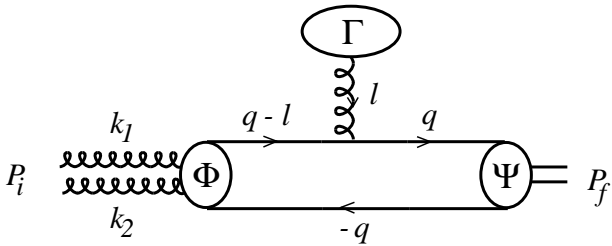}
\medskip
\caption{A perturbative interaction between the quark pair and a gluon
from the color field $\Gamma$ creates an overlap between the $\qpair$
wave function $\Phi$ from the $gg \to \qpair$ process with the physical
quarkonium wave function $\Psi$. There is a second diagram where the gluon
attaches to the antiquark.}
\end{figure}

The reinteraction of the quark pair with the external field $\Gamma$
constitutes the second part of the amplitude of Fig. 4. Since this is a
relatively hard `dipole' interaction the rescattering amplitude $\R$ is
again evaluated using PQCD. In the $\qpair$ rest frame the 3-momentum part of
the transfer $\ell$ dominates, $|\ell^0| \ll |\lvec| \ll m_Q$, since the
heavy quarks are nearly on-shell both before and after the rescattering.

The third part of the scattering amplitude is the quarkonium wave function 
$\Psi$. If at the time of rescattering the quark pair is still compact
compared to the size of physical quarkonium, the amplitude depends only on
the quarkonium wave function at the origin (or on its derivative). For
$S$-wave quarkonia the expansion factor $\rho$ is then irrelevant. For
$P$-wave quarkonia, on the other hand, the overlap is $\propto \rho$ since
their wave function vanishes at the origin.

The full amplitude of Fig. 4 is thus expressed as a double convolution,
\beqa
\M&=& \sum_{L_z,S_z} \langle LL_z; SS_z |JJ_z \rangle
\sum_{\llb,\ssb} \int \frac{d^3\pvec}{(2\pi)^3}
\frac{d^3\qvec}{(2\pi)^3} \nonumber \\ &\times&\Phi_\llb^{[8]}(\pvec)\
\R_{\llb,\ssb}(\pvec,\qvec)\ {\Psi_\ssb^{L_zS_z}(\qvec)}^*
\label{prodampl}
\eeqa
Explicit expressions for the various parts of this amplitude may be found in
\cite{hoypeg}. It turns out that \eq{prodampl} implies two distinct
mechanisms for $S$-wave quarkonium production, depending on which Lorentz
components of the external field $\Gamma^{\mu}$ are involved:
\begin{itemize}
\item[(i)] The creation of the $\qpair$ pair in an $S=L=0$ state,
followed by a spin-flip interaction with a transversely polarized gluon from
the color field $\Gamma$.
\item[(ii)] The creation of the pair in an $S=L=1$ state, followed by a
spin-conserving interaction with a longitudinally polarized gluon from the
color field $\Gamma$.
\end{itemize}
Case (i) leads to production of unpolarized $\jpsi$'s, since information
about the collision axis is not transmitted by the $S=L=0$ quark pair, and we
assume $\Gamma$ to be isotropic. On the other hand, mechanism (ii) gives
rise to transversely polarized $\jpsi$'s. 

Experimentally \cite{badier,poln,heinrich} both the $\jpsi$ and the $\psi'$
are unpolarized (at the moderate values of $x_F$ which are relevant for our
scenario). We conclude that our model correctly describes the $\psi'$
polarization data provided the rescattering involves transverse gluons
(which implies $\Gamma^0 = 0$). Conclusions for the $\jpsi$ are complicated
by the fact that about 40\% of the measured $\jpsi$ cross section stems from
$\chi_{c1,2} \to \jpsi +
\gamma$ radiative decays \cite{e705,e672}. This component has not been
removed in the polarization measurements. Hence we need to take into account
also the production and polarization of $P$-wave quarkonia.

\begin{table}[htb]
\begin{tabular}{ccccccccc}
{\LARGE\strut} & {\large $^3P_0$} & \hspace{.2cm} &
\multicolumn{2}{c}{{\large $^3P_1$}} &
\hspace{.2cm} & \multicolumn{3}{c}{{\large $^3P_2$}} \\
{\LARGE\strut} {\large $J_z$} & {\large 0} && {\large 1} &{\large 0} &&
{\large 2} & {\large 1} & {\large 0} \\
\hline
{\LARGE\strut} {\large $\frac{\sigma(^3P_J,J_z)}
{\sigma_{dir}(^3S_1)}$} & {\large $r$} && {\large $\frac{3}{2}r$} &
{\large 0} && {\large 0}  & {\large
$\frac{3}{2}r$} & {\large $2r$} \\
{\LARGE\strut} {\large $\lambda_{indir}(^3S_1)$} & {\large 0} &&
{\large $-\frac{1}{3}$} & {\large 1} && {\large 1} &
{\large $-\frac{1}{3}$} & {\large $-\frac{3}{5}$} \\
\end{tabular}
\caption{Relative $^3P_J$ cross sections and induced $J/\psi$
polarizations.}
\label{tab2}
\end{table}

It turns out that in our model the $P$-wave cross sections have the same
dependence on the external field $\Gamma$ as the $S$-wave ones. The
cross section ratios are given in Table I in terms of
\beq
r= \frac{3}{5} \rho^2 \left(\frac{R_1'/m}{R_0}\right)^2
\simeq \left\{
\begin{array}{dcc}
2.5 \, 10^{-2}\, \rho^2 & \hspace{.3cm} & (c\bar c) \\
6.5 \, 10^{-3}\, \rho^2 & \hspace{.3cm} & (b\bar b) \\
\end{array} \right.
\label{rparam}
\eeq
which depends on the expansion paramter $\rho$ of the quark pair wave
function. 

The total cross section ratio $\sigma(\chi_1):\sigma(\chi_2) = 3:5$
of Table \ref{tab2} is consistent with the experimental number $0.6 \pm
0.3$. It should be kept in mind, however, that the $\chi_{c2}$ cross section
calculated in the CSM is not far below the measured one. The CSM could
well account for half of the measured $\sigma(\chi_{c2})$. Adding this
contribution to ours will correspondingly decrease the calculated
$\sigma(\chi_1):\sigma(\chi_2)$ ratio.

The measured $\sigma(\chi_{c2})$ is somewhat larger than
$\sigma_{dir}(\jpsi)$. This is generically difficult to understand in an
approach like ours, which couples a compact quark pair directly to the
quarkonium wave function. As seen from \eq{rparam}, the $P$-wave cross
section is suppressed due to the wave function vanishing at the origin. We
need to have a considerable expansion of the quark pair wave function
between its creation and reinteraction, $\rho \sim 3$, to describe the data.
Such a large expansion makes the $\cpair$ radius comparable to that of the
$\jpsi$, \ie, the quark pair is no longer `compact'.

In Table \ref{tab2} we also show the induced $\jpsi$ polarizations from the
radiative decays of the various $P$-wave states. The $\jpsi$ polarization is
parametrized through the ratio $\lambda$ defined by
\beq
\lambda = \frac{\sigma(J_z=+1)-\sigma(J_z=0)}
{\sigma(J_z=+1)+\sigma(J_z=0)}  \label{lamdef}
\eeq
The $P$-wave helicity states which are produced in  our model induce a
longitudinal $(\lambda < 0)$ polarization for the $\jpsi$. On the other
hand, the CSM produces $\chi_{c2}$'s with $J_z = \pm 2$ \cite{vhbt}, giving 
transversely polarized $\jpsi$'s. If the CSM contributes $\sim 50\%$ of the
measured $\sigma(\chi_{c2})$, the total indirectly produced  $\jpsi$'s turn
out to be unpolarized. Since in our mechanism the directly produced
$\jpsi$ component is also unpolarized, the experimentally observed
non-polarization of the $\jpsi$ can thus be understood.

\section{Conclusions} \label{sec5}

Quarkonium cross sections are sensitive to rescattering between the
produced heavy quarks and accompanying colored partons. The standard QCD
factorization theorem which allows one to ignore such effects in hard
inclusive processes does not apply to quarkonium production. The
observed angular (de)correlations between hadroproduced $D\bar D$ pairs show
that multiple scattering effects are strong.

The quarkonium data as well as theoretical arguments suggest \cite{hoypeg} a
rescattering scenario where the relevant rescattering (\ie, the one which
affects the cross section) is perturbatively hard and occurs soon after the
heavy quarks are produced. The rescattering is associated with the
gluon-gluon fusion process and thus depends on the colliding hadrons only
via their universal gluon distributions. There is no rescattering in the
photon-gluon fusion process. 

Such a scenario is consistent with the phenomenological successes of the
Color Evaporation Model (CEM) \cite{cem1,cem2}. It avoids a theoretical
inconsistency of the CEM, namely that heavy quark spins would have to be
flipped by soft interactions. It also avoids some phenomenological problems
of the CEM, namely that radiative $\chi_c$ decays are observed to
contribute less in $\jpsi$ photoproduction than in hadroproduction
\cite{na14}, and that quarkonium hadroproduction on nuclear targets is
suppressed. Finally, the present scenario explains why the ratio
$\sigma(\psi')/\sigma_{dir}(\jpsi)$ is close to the ratio of the wave
functions at the origin, which is a `coincidence' in the CEM.

The scenario discussed above is theoretically consistent with the Color Octet
Mechanism (COM) \cite{com}, which is based on an NRQCD \cite{nrqcd}
expansion in powers of the bound state velocity $v$. We only discussed the
lowest order contribution in $v$, but added rescattering effects which there
appears to be no reason to exclude in a COM approach. From a
phenomenological point of view the scenario presented in \cite{hoypeg} is
quite different from the COM, in that the numerically important effects are
ascribed to hard rescattering rather than to relativistic bound state
corrections.

A simple and explicit model for quarkonium production according to our
scenario gave some encouraging results that were not anticipated, in
particular concerning the (non-)polarization of the $\jpsi$ and the
$\sigma(\chi_{c1})/\sigma(\chi_{c2})$ ratio. However, the model parameter
$\rho$, which describes the spatial expansion of the heavy quark pair
between its creation and rescattering times, had to be rather large
($\rho \sim 3$) to explain the measured $\sigma(\chi_{cJ})/\sigma(\jpsi)$
ratio.

The hard rescattering scenario can also be applied to quarkonium production
at large $\ptr$, where the subprocess is gluon fragmentation to heavy
quarks. It would be useful to make a quantitative study of this process in
the framework of an explicit model.

The scenario discussed above does not address the special features
of quarkonium production observed at large $x_F$ (increased nuclear
suppression and longitudinal $\jpsi$ polarization). Other effects, \eg, as
discussed in Ref. \cite{bhmt}, need to be taken into account in that
kinematic region.

\section*{Acknowledgments}
This talk is based on work done together with S. Peigne. I am grateful also 
for helpful discussions with S. J. Brodsky and A. Khodjamirian.

\end{document}